\documentclass[prd,eqsecnum,showpacs,aps]{revtex4}
\usepackage{latexsym}
\begin{document}
\title{Second-order power spectra of CMB anisotropies due to
primordial\\ random 
perturbations in flat cosmological models
}
\author{Kenji Tomita}
\affiliation{Yukawa Institute for Theoretical Physics, 
Kyoto University, Kyoto 606-8502, Japan}
\date{\today}

\begin{abstract}
Second-order power spectra of Cosmic Microwave Background (CMB)
anisotropies due to random primordial perturbations at the matter
dominant stage are studied, 
based on the relativistic second-order theory of perturbations in
 flat cosmological models and on the
second-order formula of CMB anisotropies derived by Mollerach and
Matarrese. So far the second-order integrated Sachs-Wolfe effect has
been analyzed using the three-point correlation or bispectrum. In this
paper we derive the second-order term of power spectra given
using the two-point correlation of temperature fluctuations. 

The second-order density perturbations are small,
compared with the first-order ones. The second-order power spectra of
CMB anisotropies, however, are not small at all, compared with the
first-order power spectra, because at the early stage the first-order
integrated Sachs-Wolfe effect is very small and the second-order
integrated Sachs-Wolfe effect may be dominant over the first-order
ones. So their characteristic behaviors may be measured
through the future precise observation and bring useful informations
on the structure and evolution of our universe in the future. 
\end{abstract}

\maketitle


\section{Introduction}
\label{sec:level1}

In most studies of Cosmic Microwave Background (CMB) anisotropies, the
comparison between observed and theoretical quantities have so far
been done, assuming the linear approximation for cosmological
perturbations. It seems to be successful enough to determine the
cosmological parameters \cite{map,spg,komt}. The present state of our
universe is, however, locally complicated and associated with
nonlinear behavior on various 
scales, and so the observed quantities of CMB anisotropies may include
some small effects caused through nonlinear process by various
primordial perturbations. So far the nonlinearity in CMB anisotropies
has been treated as the second-order integrated Sachs-Wolfe effect,
which was analyzed using the three-point correlation and
bispectrum.\cite{mglm,munshi,sg} Recently a general treatment of
second-order temperature fluctuations has been systematically studied
by Bartolo, Matarrese and Riotto\cite{bart1,bart2} due to the
transfer function which was derived from the Boltzmann equation, so as
to be 
applied not only to the integrated Sachs-Wolfe effect at the
matter-dominant stage, but also the nonlinear effect at the
recombination epoch and the primordial nonlinear effect.  

In recent years, on the other hand, we have studied these nonlinear
effects of inhomogeneities on CMB anisotropies at the matter-dominant
stage, based on the relativistic  
second-order theory of cosmological perturbations,\cite{tom1} and on
Mollerach and Matarrese's 
second-order formula of CMB anisotropies\cite{cmb}. In previous
papers,\cite{tom2,tom3,tom4,is1,is2} we have studied the 
second-order effects of a local special large-scale inhomogeneity 
on CMB anisotropies, paying attention to the interaction between it
 and primordial random perturbations. In this relativistic theory we
used the first-order and second-order perturbations in pressureless
matter which were derived in compact and analytic forms. 
In the present paper we study the
nonlinear effect of only primordial random perturbations on CMB
anisotropies and derive the second-order power spectra, which is given
as the second term in the two-point correlation of CMB
anisotropies. This is a nonlinear correction to the first-order power
spectra and  
different from the three-point correlation and bispectrum which have
been used in the above other works to investigate uniquely the
non-Gaussianity in the perturbations generated in various stages. 

The second-order random density perturbations themselves are small, 
compared with the first-order ones. In the $\Lambda$-dominated model,
however, the second-order power spectra of  
CMB anisotropies are not small at all, compared with the
first-order power spectra, because at the early stage the first-order
integrated Sachs-Wolfe effect (ISW) is very small and the second-order
ISW may be dominant over the first-order one.
Especially in the local void model with $\Lambda =
0$\cite{lvm,cele,aln,bis}, the exterior region is described by use of 
the Einstein-de Sitter model, in which the first-order ISW vanishes 
and the second-order ISW is indispensable.
So their characteristic behaviors may be measured
through the future precise observation and bring useful informations
on the structure and evolution of our universe in the future.

In Sec. II, we show the second-order perturbations in 
general flat cosmological models  and the
corresponding CMB anisotropies.  In Sec. III, we derive the expressions
for the second-order power spectra of CMB anisotropies,
due to primordial random perturbations. Concluding remarks follow in
Section IV. 
 
\section{Second-order metric perturbations and temperature anisotropies}
\label{sec:level2}

First we review the background spacetime and the perturbations which
were derived in the previous paper\cite{tom2}. The background flat
model with dust matter is expressed as 
\begin{equation}
  \label{eq:m1}
 ds^2 =  g_{\mu\nu} dx^\mu dx^\nu = 
a^2 (\eta) [- d \eta^2 + \delta_{ij} dx^i dx^j],
\end{equation}
where the Greek and Latin letters denote $0,1,2,3$ and $1,2,3$,
respectively, and
$\delta_{ij} (= \delta^i_j = \delta^{ij})$ are the Kronecker
delta. The conformal time $\eta (=x^0)$ is related to the cosmic time
$t$ by $dt = a(\eta) d\eta$. The matter density $\rho$ and the scale
factor $a$ have the relations
\begin{equation}
  \label{eq:m2}
\rho a^2 = 3(a'/a)^2 - \Lambda a^2, \quad {\rm and } \quad 
\rho a^3 = \rho_0,
\end{equation}
where a prime denotes $\partial/\partial \eta$,  $\Lambda$ is the
cosmological constant,  $\rho_0$ is an integration constant and the
units $8\pi G = c =1$ are used.

The first-order and second-order metric perturbations
$\mathop{\delta}_1 g_{\mu\nu} (\equiv 
h_{\mu\nu})$ and $\mathop{\delta}_2 g_{\mu\nu} (\equiv
\ell_{\mu\nu})$, respectively, were derived explicitly by imposing the
synchronous coordinate condition:
\begin{equation}
  \label{eq:m3}  
h_{00} = h_{0i} = 0 \quad {\rm and} \quad \ell_{00} = \ell_{0i} = 0.
\end{equation}
Here we show their expressions only in the growing mode:
\begin{eqnarray}
  \label{eq:m4}
h^j_i &=&  P(\eta) F_{,ij}, \cr
\ell^j_i &=& P(\eta) L^j_i + P^2 (\eta) M^j_i + Q(\eta) N^{|j}_{|i} +
C^j_i, 
\end{eqnarray} 
where $F$ is an arbitrary potential function of spatial coordinates
$x^1, x^2$ and $x^3, \ h^j_i = \delta^{jl}h_{li}, \ N^{|j}_{|i} =
\delta^{jl} N_{|li} = N_{,ij}, \ F_{,ij}$ means $\partial^2 F/\partial
x^i \partial x^j,$ and $P(\eta)$ and $Q(\eta)$ satisfy 
\begin{eqnarray}
  \label{eq:m5}
P'' &+& {2a' \over a} P' -1 = 0, \cr
Q'' &+& {2a' \over a} Q' = -\Bigl[P - {5 \over 2} (P')^2\Bigr].
\end{eqnarray}
The three-dimensional covariant derivative $|i$ are defined in the
space with metric $dl^2 = \delta_{ij} dx^i dx^j$ and their suffixes
are raised and lowered by use of $\delta_{ij}$. 
The functions $L^j_i$ and $M^j_i$ are defined by
\begin{eqnarray}
  \label{eq:m6}
L^j_i &=& {1 \over 2}\Bigl[-3 F_{,i} F_{,j} -2 F \cdot F_{,ij} + {1 \over 2}
\delta_{ij} F_{,l} F_{,l}\Bigr], \cr
M^j_i &=& {1 \over 28}\Big\{19F_{,il} F_{,jl} - 12 F_{,ij} \Delta F -
3\delta_{ij} \Bigl[F_{,kl} F_{,kl} -(\Delta F)^2 \Bigr]\Big\}
\end{eqnarray} 
and $N$ is defined by
\begin{equation}
  \label{eq:m7}
\Delta N = {1 \over 28} \Bigl[(\Delta F)^2 - F_{,kl}F_{,kl}\Bigr].
\end{equation} 
The last term $C^l_i$ satisfies the wave equation
\begin{equation}
  \label{eq:m8}
\Box C^j_i = {3 \over 14}(P/a)^2 G^j_i + {1 \over 7}\Bigl[P - {5 \over 2}
(P')^2 \Bigr] \tilde{G}^j_i,
\end{equation} 
where $G^j_i$ and $\tilde{G}^j_i$ are second-order traceless and transverse
functions of spatial coordinates, and the operator $\Box$ is defined by
\begin{equation}
  \label{eq:m9}
\Box \phi \equiv g^{\mu\nu} \phi_{;\mu\nu} = -a^{-2}
\Bigl(\partial^2/\partial\eta^2 + {2a' \over a}\partial/\partial \eta -
\Delta \Bigr) \phi, 
\end{equation} 
where $;$ denotes the four-dimensional covariant derivative.
So $C^j_i$ represents the second-order gravitational waves caused by
the first-order density perturbations. 

The velocity perturbations $\mathop{\delta}_1 u^\mu$ and
$\mathop{\delta}_2 u^\mu$ vanish, i.e. \ $\mathop{\delta}_1 u^0 =
\mathop{\delta}_1 u^i =0$ and $\mathop{\delta}_2 u^0 =
\mathop{\delta}_2 u^i =0$, and the density perturbations are 
\begin{eqnarray}
  \label{eq:m10}
\mathop{\delta}_1 \rho/\rho &=& {1 \over \rho a^2} \Bigl({a'\over a}P'
-1\Bigr) \Delta F, \cr
\mathop{\delta}_2 \rho/\rho &=& {1 \over 2\rho a^2}\Bigl\{{1 \over
2}(1 - {a' \over a}P') (3F_{,l}F_{,l} + 8F\Delta F) +{1 \over 2}P
[(\Delta F)^2 + F_{,kl}F_{,kl}] \cr
 &+& {1 \over 4}\Bigl[(P')^2 - {2 \over 7}{a'\over a}Q'\Bigr] [(\Delta F)^2 -
F_{,kl}F_{,kl}] - {1 \over 7} {a'\over a}PP' [4F_{,kl}F_{,kl} +
3(\Delta F)^2]  \Big\}.
\end{eqnarray} 

Next let us consider the CMB temperature $T = T^{(0)} (1 + \delta
T/T)$, in which $T^{(0)}$ is the background temperature and $\delta T/T
(= \mathop{\delta}_1 T/T + \mathop{\delta}_2 T/T)$ is the
perturbations. The 
present temperature $T^{(0)}_o$ is related to the emitted background
temperature $T^{(0)}_e$ at the recombination epoch by $T^{(0)}_e = (1+
z_e) T^{(0)}_o$, the temperature perturbation $\tau \equiv (\delta
T/T)_e$ at the recombination epoch is determined by the physical state
before that epoch, and the present temperature perturbations $(\delta
T/T)_o$ is related to $(\delta T/T)_e$ by the gravitational
perturbations along the light ray from the recombination epoch to the 
present epoch. The light ray in the unperturbed state is described 
using the background wave vector 
$k^\mu \ (\equiv dx^\mu/d \lambda)$, where $\lambda$ is the affine
parameter, and its component is $k^{(0)\mu} = (1, -e^i)$, and the ray
is given by $x^{(0)\mu} = [\lambda, (\lambda_0 - \lambda) e^i]$, where
$e^i$ is the directional unit vector. Here and in the following the
suffixes $o$ and $e$ for $\lambda, \eta$ and $r$ denote the present
(observed) epoch and the recombination (emitted) epoch, respectively.

The first-order temperature perturbation is 
\begin{equation}
  \label{eq:m11}
\mathop{\delta}_1 T/T =  \tau + {1 \over 2}\int^{\lambda_e}_{\lambda_o}
d\lambda P'(\eta) F_{,ij} e^i e^j.
\end{equation}
Using the relations $dP/d\lambda = P'$ and $dF/d\lambda = - F_{,i}
e^i$, this equation (\ref{eq:m11}) is expressed as
\begin{equation}
  \label{eq:m12}
\mathop{\delta}_1 T/T =  \Theta_1 +\Theta_2
\end{equation}
where 
\begin{eqnarray}
  \label{eq:m13}
\Theta_1 &\equiv& \tau - {1 \over 2}[(P' F_{,i})_e - (P' F_{,i})_o]
e^i, \cr 
\Theta_2 &\equiv& {1 \over 2}\int^{\lambda_e}_{\lambda_o}
d\lambda P''(\eta) F_{,i} e^i.
\end{eqnarray}
$\Theta_1$ and $\Theta_2$ represent the intrinsic and Sachs-Wolfe
effects, respectively. The latter can be divided into the ordinary
Sachs-Wolfe effect $\Theta_{sac}$ and the Integrated Sachs-Wolfe effect
$\Theta_{isw}$, where 
\begin{eqnarray}
  \label{eq:m14}
\Theta_{sac} &\equiv& {1 \over 2} [(P''F)_e - (P''F)_o], \cr
\Theta_{isw} &\equiv& {1 \over 2}\int^{\lambda_e}_{\lambda_o}
d\lambda P'''(\eta) F.
\end{eqnarray}
The second-order temperature perturbation is 
\begin{eqnarray}
  \label{eq:m15}
\mathop{\delta}_2 T/T &=& I_1 (\lambda_e) \Bigl[{1 \over 2} I_1 (\lambda_e)
- \tau\Bigr] - [{A^{(1)}_e}' + \tau_{,i} e^i] \int^{\lambda_e}_{\lambda_o}
d{\lambda} A^{(1)} \cr
&-& \int^{\lambda_e}_{\lambda_o} d{\lambda} \Big\{{1 \over 2} {A^{(2)}}' 
+ A^{(1)}{A^{(1)}}' - {A^{(1)}}'' 
\int^{\lambda}_{\lambda_o} d\bar{\lambda} A^{(1)}(\bar{\lambda})
\Big\} + {\partial\tau \over \partial d^i} d^{(1)i},
\end{eqnarray}
where $(\eta, x^i) = (\lambda, \lambda_o - \lambda)$ in the 
integrands and
\begin{eqnarray}
  \label{eq:m16}
I_1 (\lambda_e) &=& - {1 \over 2} \int^{\lambda_e}_{\lambda_o}
d{\lambda} P' F_{,ij} e^i e^j, \cr
A^{(1)} &=& - {1 \over 2}  P F_{,ij} e^i e^j , \cr
A^{(2)} &=& - {1 \over 2} [P L^j_i + P^2 M^j_i + Q N_{,ij} +
C^j_i ] e^i e^j. 
\end{eqnarray}
These expressions were derived in \S 3 of a previous paper.\cite{tom1}

\section{Power spectra of second-order CMB anisotropies}
\label{sec:level3}

We consider primordial scalar perturbations with $F$ defined by
\begin{equation}
  \label{eq:c1}
F = \int d{\bf k} \alpha ({\bf k}) e^{i{\bf kx}}, 
\end{equation}
where the spatial average for $\alpha ({\bf k})$ is given by
\begin{equation}
  \label{eq:c2}
\langle \alpha ({\bf k}) \alpha ({\bf \bar{k}}) \rangle = (2\pi)^{-2}
{\cal P}_F ({\bf k}) \delta ({\bf k} + {\bf \bar{k}}),
\end{equation}
with 
\begin{equation}
  \label{eq:c3}
{\cal P}_F ({\bf k}) = {\cal P}_{F0} k^{-3} (k/k_0)^{n-1} T_s^2 (k),
\end{equation}
where $T_s (k)$ is the matter transfer function \cite{sugi} and ${\cal
P}_{F0}$ is the normalization constant. 

Then the first-order temperature perturbations are 
\begin{equation}
  \label{eq:c4}
\mathop{\delta}_1 T/T \equiv \Theta_P = -{1 \over 2} \int
d{\bf k} \alpha ({\bf k}) 
\int^{\lambda_e}_{\lambda_o} d\lambda P' (\eta) (k\mu)^2 e^{i{\bf
kx}},
\end{equation}
where ${\bf x} = r{\bf e}, {\bf kx} = kr\mu, \mu \equiv \cos
\theta_k$ and $\theta_k$ is the angle between the wave vector $k^i$
and a unit vector $e^i$.
This equation can be rewritten as
\begin{equation}
  \label{eq:c5}
\Theta_P = \int d{\bf k} \alpha ({\bf k}) \Bigl\{- {1 \over 2}[(P'')_o +
ik (P')_o P_1(\mu)] + {1 \over 2} \sum_l (-i)^l (2l+1) {\Theta}_{P(l)}
P_l (\mu) \Bigr\},  
\end{equation}
where $P_l (\mu)$ is the Legendre polynomial and
\begin{equation}
  \label{eq:c6}
{\Theta}_{P(l)} = \int^{\lambda_e}_{\lambda_o} d\lambda  P'''
(\eta) j_l(kr) - \{k(P')_e (2l+1)^{-1}[(l+1) j_{l+1} (kr_e) - l
j_{l-1}(kr_e)] + j_l (kr_e) (P'')_e \}.
\end{equation}
In these equations, we have $\eta = \lambda$ and $r = \lambda_o -
\lambda$. In the derivation of Eq.(\ref{eq:c6}), we used the
relations\cite{zal,hu}  
\begin{equation}
  \label{eq:c7}
e^{i{\bf kx}} = e^{ikr\mu} = \sum_l (-i)^l (2l+1) j_l (kr)
P_l (\mu) 
\end{equation}
and
\begin{equation}
  \label{eq:c8}
(2l +1) \mu P_l(\mu) = (l+1) P_{l+1}(\mu) + l P_{l-1}(\mu).
\end{equation}
The components of the unit vector $e^i$ are expressed as $e^1 = \sin
\theta \cos \phi, 
e^2 = \sin \theta \sin \phi, e^3 = \cos \theta$ with respect to $x^1,
x^2, x^3$ axes, respectively.

In order to derive the power spectra, we take
the statistical average $\langle \rangle$ for the primordial
perturbations, and $\langle (\delta T/T)^2 
\rangle$ is expressed for the first-order anisotropies as
\begin{equation}
  \label{eq:c9}
\langle (\mathop{\delta}_1 T/T)^2 \rangle = \langle (\Theta_P)^2
\rangle = (T_0)^{-2}  \sum_l {2l+1 \over 4\pi} C_l.
\end{equation}
The power spectra $C_l$ are
\begin{equation}
  \label{eq:c10}
C_l =  (T_0)^2 \int dk k^2 {\cal P}_F (k)
|{\cal H}_P^{(l)} (k) |^2, 
\end{equation}
where $T_0$ is the present CMB temperature and
\begin{eqnarray}
  \label{eq:c10a} 
{\cal H}_P^{(0)} (k) &=& -(P'')_o - k(P')_e j_1(kr_e) - (P'')_e j_0(kr_e)
+ \int^{\lambda_e}_{\lambda_o} d\lambda P''' j_0(kr), \cr
{\cal H}_P^{(1)} (k) &=& {1 \over 3}k[(P')_o - (P')_e] j_1^{(1)}(kr_e)
-(P'')_e j_1 (kr_e) + \int^{\lambda_e}_{\lambda_o} d\lambda P''' j_1
(kr). 
\end{eqnarray}
For $l \geq 2$, we have
\begin{equation}
  \label{eq:c11}
{\cal H}_P^{(l)} (k) = k (P')_e j_l^{(1)} (kr_e) - (P'')_e j_l (kr_e) +
\int^{\lambda_e}_{\lambda_o} d\lambda P''' j_l (kr). 
\end{equation}
In the derivation of $C_l$, we used Eq.(\ref{eq:c2}) for $\langle
\alpha ({\bf k}) \alpha ({\bf \bar{k}}) \rangle$ and a mathematical
formula for $P_l (\mu)$: 
\begin{equation}
  \label{eq:c11a}
\ \int^1_{-1} P_l (\mu) P_{l'} (\mu) d\mu = \ 2/(2l+1), \ 0 \quad
{\rm for} \quad  l=l', \ l \ne l', 
\end{equation}
respectively.
For the two directions with unit vectors ${\bf e}_{1}$ and ${\bf
e}_{2}$, we have the correlation                 
\begin{equation}
  \label{eq:c12}
(T_0)^{2} \langle \Theta_P({\bf e}_{1}) \Theta_P({\bf e}_{2}) \rangle
=  \sum_l {2l+1 \over 4\pi} C_l P_l (\cos \beta). 
\end{equation}
where the product ${\bf e}_{1} {\bf e}_{2}$ is equal to $\cos \beta$.

For the second-order temperature anisotropies, we obtain from
Eqs.(\ref{eq:m15}), (\ref{eq:m16}) and (\ref{eq:c1}) 
\begin{eqnarray}
  \label{eq:c13}
\mathop{\delta}_2 T/T &=& \int\int d{\bf k}d\bar{\bf k} \alpha
({\bf k}) \alpha (\bar{\bf k}) \Bigl\{{1\over 8}
\int^{\lambda_e}_{\lambda_o}  
\int^{\lambda_e}_{\lambda_o}  d\lambda d\bar{\lambda}
P'(\lambda) P'(\bar{\lambda}) (k_e \bar{k}_e)^2 e^{i({\bf k}
{\bf x}+ \bar{\bf k} \bar{\bf x})} \cr  
&+& \int^{\lambda_e}_{\lambda_o} d\lambda \Bigl[{1\over 8}
P'(\lambda) \Bigl(3k_e \bar{k}_e +(k_e)^2 +(\bar{k}_e)^2  - {1\over
2}{\bf k} \bar{\bf k}\Bigr) \cr
&+& {1\over 56}  
P(\lambda) P'(\lambda)\ \Bigl(19k_e \bar{k}_e {\bf k}\bar{{\bf k}} -
14 (k_e \bar{k}_e)^2 -6(\bar{k}_e k)^2 -6(k_e \bar{k})^2
- 3({\bf k}\bar{\bf k})^2 + 3k^2 (\bar{k})^2
\Bigr) \cr 
&+& {1 \over 112} Q'(\lambda) (k_e+\bar{k}_e)^2 \Bigl(k^2\bar{k}^2 -
({\bf k}{\bf \bar{k}})^2\Bigr)/({\bf k}+{\bf \bar{k}})^2 \Bigr]
e^{i({\bf k} +\bar{{\bf k}}){\bf x}} \cr
&+& {1\over 4} \int^{\lambda_e}_{\lambda_o} d \lambda
\Bigl[P''(\lambda) \int^\lambda_{\lambda_o} d\bar{\lambda}
P({\bar{\lambda}}) (k_e \bar{k}_e)^2 \Bigr] e^{i({\bf k}{\bf 
x}+ \bar{\bf k} {\bf \bar{x}})} \Bigr\}, 
\end{eqnarray}
where $k_e$ stands for ${\bf k} {\bf e}, \ P'(\lambda) \equiv
dP(\lambda) 
/d\lambda$, and $P'(\bar{\lambda}) \equiv dP(\bar{\lambda})
/d\bar{\lambda}$. Here $\tau$ and ${A_e^{(1)}}'$ in Eq.(\ref{eq:m15})
were neglected, because we pay attentions to the Sachs-Wolfe effect
after the recombination 
epoch, and also the term with $C^j_i$ in Eq.(\ref{eq:m16}) was
neglected, because the  
contribution of gravitational radiation is very small. It is found
from the above equation that the
average $\langle \mathop{\delta}_2 T/T \rangle$ does not vanish in 
contrast to the vanishing first-order one ($\langle \mathop{\delta}_1 
T/T \rangle$), and we obtain 
\begin{eqnarray}
  \label{eq:c14}
\langle \mathop{\delta}_2 T/T \rangle &=& {1\over 8} \int d{\bf k}
(2\pi)^{-2} 
{\cal P}_F ({\bf k})  \Bigl\{ \int^{\lambda_e}_{\lambda_o}
\int^{\lambda_e}_{\lambda_o} d\lambda d\bar{\lambda}
P'(\lambda) P'(\bar{\lambda}) (k_e)^4 e^{i{\bf k}({\bf x}
-\bar{{\bf x}})} \cr
&+& \int^{\lambda_e}_{\lambda_o} d\lambda \Bigl[ P'(\lambda)
\Bigl( -(k_e)^2 + 
{1\over 2} k^2\Bigr) + P(\lambda) P'(\lambda) \Bigl(k^2
-2(k_e)^2\Bigr)(k_e)^2\Bigr] \cr 
&+& 2 \int^{\lambda_e}_{\lambda_o} d\lambda
P''(\lambda) \int^\lambda_{\lambda_o} d\bar{\lambda} 
P(\bar{\lambda}) (k_e)^4 e^{i{\bf k}({\bf x}
-\bar{{\bf x}})}  \Bigr\},
\end{eqnarray}
where $k_e \equiv {\bf k} {\bf e} = k\mu$ and $d{\bf k} = dk k^2 d\phi
d\mu$. Here we have the 
relations ${\bf x} = r{\bf e}, \ \bar{{\bf x}} = \bar{r}{\bf e}$, and
so ${\bf kx} = k_e r = kr \mu, \ - {\bf k\bar{x}} = k \bar{r} (-\mu)$.
Thus $e^{i{\bf k}({\bf x} -\bar{{\bf x}})}$ is expanded as
\begin{eqnarray}
  \label{eq:c15}
e^{i{\bf k}({\bf x} -\bar{{\bf x}})} &=& \sum_l (-i)^l (2l+1) j_l (kr)
P_l(\mu) \times \sum_{l'}(-i)^{l'} (2l'+1) j_{l'} (k\bar{r}) P_{l'}
(-\mu)\cr
&=& \sum_{l,l'} (-i)^{l}\ i^{l'}\ (2l+1)(2l'+1)\ j_l (kr) j_{l'}
(k\bar{r}) \ P_l(\mu) P_{l'}(\mu).
\end{eqnarray}
Here for the reduction of Eq.(\ref{eq:c14}), we derive the following
relations using Eq.(\ref{eq:c8})
\begin{equation}
  \label{eq:c16}
\sum_l j_l\ \mu P_l = \sum_l j_l^{(1)} P_l, \qquad
\sum_l j_l\ \mu^2 P_l = \sum_l j_l^{(2)} P_l, \qquad {\rm and} \qquad
\sum_l j_l\ \mu^3 P_l = \sum_l j_l^{(3)} P_l,
\end{equation}
where
\begin{eqnarray}
  \label{eq:c17}
j_l^{(1)}(kr) &\equiv& {l \over 2l-1} j_{l-1} (kr) + {l+1 \over 2l+3}
j_{l+1} (kr), \cr 
j_l^{(2)}(kr) &\equiv& {(l-1)l \over (2l-3)(2l-1)}j_{l-2} (kr) +
{1\over 2l+1} \Bigl[{l^2 \over 2l-1} 
+ {(l+1)^2\over (2l+3)}\Bigr] j_l
(kr) + {(l+1)(l+2) \over (2l+3)(2l+5)} j_{l+2} (kr), \cr
j_l^{(3)}(kr) &\equiv& {(l-2)(l-1)l \over (2l-5)(2l-3)(2l-1)} j_{l-3}
(kr) + {l\over 2l-1} \Bigl[{(l-1)^2 \over (2l-3)(2l-1)}\cr
&+& {l^2\over (2l-1)(2l+1)} +{(l+1)^2 \over (2l+1)(2l+3)} \Bigr]
j_{l-1} (kr) + {l+1 \over 2l+3} \Bigl[{l^2 \over (2l-1)(2l+1)}\cr
&+& {(l+1)^2\over (2l+1)(2l+3)} +{(l+2)^2 \over (2l+3)(2l+5)} \Bigr]
j_{l+1} (kr) 
+  {(l+1)(l+2)(l+3) \over (2l+3)(2l+5)(2l+7)} j_{l+3} (kr).
\end{eqnarray}
Then we have
\begin{equation}
  \label{eq:c18}
\mu^4 e^{i{\bf k}({\bf x} -\bar{{\bf x}})} = \sum_{l,l'} (-i)^l
i^{l'} (2l+1) (2l'+1) j_l^{(2)} (kr) j_{l'}^{(2)} (k\bar{r}) P_l(\mu)
P_{l'} (\mu). 
\end{equation}
Using these relations and executing the integrations in
Eq.(\ref{eq:c14}) with respect to $\phi$ and $\mu$, we obtain 
\begin{eqnarray}
  \label{eq:c19}
\langle \mathop{\delta}_2 T/T \rangle &=&  (2\pi)^{-1} \int dk k^2
{\cal P}_F ({\bf k})  \Bigl\{{1\over 4} k^4 \sum_l (2l+1)
\Bigl[\int^{\lambda_e}_{\lambda_o} 
d\lambda  P'(\lambda) j_l^{(2)} (kr)\Bigr]^2 \cr
&+& {1\over 24} (P_e - P_o) k^2 + {17\over 840} (P_e^2 -P_o^2)
k^4  \cr
&+& {1\over 2} k^4 \sum_l (2l+1) \int^{\lambda_e}_{\lambda_o} d\lambda
P''(\lambda) 
j_l^{(2)} (kr) \int^\lambda_{\lambda_o} d\bar{\lambda}
P(\bar{\lambda}) j_l^{(2)}(k\bar{r}) \Bigr\}, 
\end{eqnarray}
where we used the relation (\ref{eq:c11a}), and $P_e = P(\lambda_e)$ and
$P_o = P(\lambda_o)$. 

When we take into account also the third-order anisotropies
$\mathop{\delta}_3 T/T$, we have 
\begin{eqnarray}
  \label{eq:c20a}
\mathop{\delta} T/T &=& \mathop{\delta}_1 T/T + \mathop{\delta}_2 T/T
+ \mathop{\delta}_3 T/T \cr
&=& \mathop{\delta}_1 T/T + \langle \mathop{\delta}_2 T/T \rangle
+ \Theta_{pp} + \mathop{\delta}_3 T/T, 
\end{eqnarray}
where  $\Theta_{pp} \equiv  \mathop{\delta}_2 T/T - \langle
\mathop{\delta}_2 T/T \rangle$.
Then the total average of $(\delta T/T)^2$ is expressed as
\begin{equation}
  \label{eq:c21}
\langle (\mathop{\delta} T/T)^2 \rangle =
 \langle \Theta_p^2 \rangle + (\langle \mathop{\delta}_2 T/T
\rangle)^2 + \langle \Theta_{pp}^2 \rangle + 2 \langle
\mathop{\delta}_1 T/T \mathop{\delta}_3 T/T \rangle.
\end{equation}
The last term is of the same order as the second and third terms.
At present we have not obtained any concrete expression of third-order
metric perturbations yet and so the last term is not analyzed, 
while the formula for $\mathop{\delta}_3 T/T$ and formal solutions 
have recently been derived by
D'Amico et al.\cite{dam} and the perturbative equations to the third
order were studied by Hwang and Noh\cite{hwang} with respect to the
relativistic-Newtonian correspondence. 
 Here $\langle \mathop{\delta}_2 T/T
\rangle$ is the monopole component without angular dependence and 
$\Theta_{pp}$ is the renormalized second-order temperature
fluctuation, which have been discussed by Munshi et al.\cite{munshi}

Now let us make a reduction of $\langle \Theta_{pp}^2 \rangle$. It is
expressed as 
\begin{equation}
  \label{eq:c22}
\langle \Theta_{pp}^2 \rangle = \langle \Bigl(\mathop{\delta}_2
T/T(\alpha({\bf k})\alpha(\bar{\bf k})) - \langle \mathop{\delta}_2 T/T
\rangle \Bigr) \Bigl(\mathop{\delta}_2 T/T(\alpha(\bar{\bar{\bf
k}})\alpha(\bar{\bar{\bar{\bf k}}})) -  \langle \mathop{\delta}_2 T/T
\rangle \Bigr) \rangle, 
\end{equation}
where $\mathop{\delta}_2 T/T(\alpha({\bf k})\alpha(\bar{\bf k}))$ is
given 
by Eq.(\ref{eq:c13}) and $\mathop{\delta}_2 T/T(\alpha(\bar{\bar{\bf
k}})\alpha(\bar{\bar{\bar{\bf k}}}))$ is obtained from
Eq.(\ref{eq:c13}) by 
replacing ${\bf k}, \bar{\bf k}$ by $ \bar{\bar{\bf k}},\
 \bar{\bar{\bar{\bf k}}}$. The averaging process in Eq.(\ref{eq:c22})
is performed in the two sets: 
$\langle \alpha({\bf k})\alpha(\bar{\bar{\bf k}}) \rangle, \langle 
\alpha(\bar{\bf k})\alpha(\bar{\bar{\bar{\bf k}}}) \rangle$ and
$\langle \alpha({\bf  
k})\alpha(\bar{\bar{\bar{\bf k}}}) \rangle, \langle \alpha(\bar{\bf k})
\alpha(\bar{\bar{\bf k}}) \rangle$, and the average process for
$\langle \alpha({\bf k})\alpha(\bar{\bf k}) \rangle$ and $\langle 
\alpha(\bar{\bar{\bf k}})\alpha(\bar{\bar{\bar{\bf k}}}) \rangle$ is
excluded by subtracting $\langle \mathop{\delta}_2 T/T
\rangle$ from $\mathop{\delta}_2 T/T(\alpha({\bf k})\alpha(\bar{\bf
k}))$ and $\mathop{\delta}_2 T/T(\alpha(\bar{\bar{\bf 
k}})\alpha(\bar{\bar{\bar{\bf k}}}))$.  
When we sum these average values in the above two sets
and symmetrize the expression with respect to ${\bf k}$ and $\bar{\bf
k}$, we obtain from Eq.(\ref{eq:c22})   
\begin{eqnarray}
  \label{eq:c23}
\langle \Theta_{pp}^2 \rangle &=& 2 \int\int dk d\bar{k} (k\bar{k})^2
d\phi_k d\bar{\phi}_k d\mu d\bar{\mu} (2\pi)^{-4}{\cal P}_F(k){\cal
P}_F(\bar{k})\cr
&\times& \Bigl\{{1\over 8} \int^{\lambda_e}_{\lambda_o}
\int^{\lambda_e}_{\lambda_o} d\lambda d\bar{\lambda} P'(\lambda) 
P'(\bar{\lambda}) (k_e \bar{k}_e)^2 e^{i({\bf kx} +{\bf
\bar{k}\bar{x}})} 
+ \int^{\lambda_e}_{\lambda_o} d\lambda \Bigl[{1\over 8} P'(\lambda)
\Bigl(3k_e\bar{k}_e +(k_e)^2+ (\bar{k}_e)^2 -{1\over 2}{\bf
k\bar{k}}\Bigr) \cr
&+& {1\over 56} P(\lambda) P'(\lambda) 
\Bigl(19k_e \bar{k}_e {\bf k\bar{k}} - 14(k_e \bar{k}_e)^2 
-6(\bar{k}_e k)^2 -6(k_e \bar{k})^2 
- 3({\bf k\bar{k}})^2+ 3k^2 (\bar{k})^2\Bigr) \Bigr] e^{i({\bf k}
+{\bf \bar{k}}){\bf x}}     \cr
&+& {1\over 4} \int^{\lambda_e}_{\lambda_o} d\lambda P''(\lambda)
 \int^\lambda_{\lambda_o} 
 P(\bar{\lambda}) (k_e \bar{k}_e)^2 e^{i({\bf kx} +{\bf
\bar{k}\bar{x}})} \Bigr\}\cr
&\times& \Bigl\{{1\over 8} \int^{\lambda_e}_{\lambda_o}
\int^{\lambda_e}_{\lambda_o} d\eta d\bar{\eta} P'(\eta) 
P'(\bar{\eta}) (k_e \bar{k}_e)^2 e^{-i({\bf k y} +{\bf
\bar{k}\bar{y}})} + \int^{\lambda_e}_{\lambda_o} d\eta \Bigl[{1\over
8} P'(\eta) \Bigl(3k_e\bar{k}_e +(k_e)^2+ 
(\bar{k}_e)^2 -{1\over 2}{\bf k\bar{k}}\Bigr) \cr
&+& {1\over 56} P(\eta)P'(\eta) \Bigl(19k_e \bar{k}_e {\bf k\bar{k}} 
- 14(k_e \bar{k}_e)^2 -6(\bar{k}_e k)^2 -6(k_e \bar{k})^2 - 3({\bf
k\bar{k}})^2 
+ 3k^2 (\bar{k})^2\Bigr) \Bigr] e^{-i({\bf k} +{\bf \bar{k}})
{{\bf y}}} \cr     
&+& {1\over 4} \int^{\lambda_e}_{\lambda_o} d\eta P''(\eta)
\int^\eta_{\lambda_o} d\bar{\eta} 
 P(\bar{\eta}) (k_e \bar{k}_e)^2 e^{-i({\bf k y} +{\bf
\bar{k} \bar{y}})} \Bigr\},
\end{eqnarray}
where ${\bf x, \bar{x}}$ and ${\bf y, \bar{y}}$ are functions of
$\lambda, \bar{\lambda}$ and $\eta, \bar{\eta}$, respectively, and we
neglected the terms with $Q$ and $ N^{|j}_{|i}$, 
because we have $Q/P^2 < 10^{-2}$ always and they are very small.
Especially $Q$ vanishes in the case $\Lambda = 0$.
Here, when we consider an orthonormal triad vector $e^i_{(1)},
e^i_{(2)}$, and $e^i_{(3)}\ (= e^i)$, the components of ${\bf k}$ and
${\bf \bar{k}}$ with respect to this triad are expressed as ${\bf k} = k
(\sin \theta_k \cos \phi_k, \sin \theta_k \sin \phi_k, \cos
\theta_k)$ and ${\bf \bar{k}} = \bar{k}
(\sin \bar{\theta}_k \cos \bar{\phi}_k, \sin \bar{\theta}_k \sin
\bar{\phi}_k, \cos \bar{\theta}_k)$, and so $\mu = \cos \theta_k$ and
$\bar{\mu} = \cos \bar{\theta}_k$. 

Next let us take a notice of terms with ${\bf k\bar{k}}$ and $({\bf
k\bar{k}})^2$ which can be expressed as
\begin{eqnarray}
  \label{eq:c24}   
{\bf k\bar{k}} &=& k\bar{k} \ [\sin \theta_k \sin \bar{\theta}_k \cos
(\phi_k -\bar{\phi}_k) + \cos \theta_k \cos \bar{\theta}_k], \cr
({\bf k\bar{k}})^2 &=& (k\bar{k})^2 \ \{(\cos \theta_k \cos
\bar{\theta}_k)^2 +{1\over 2} 
(\sin \theta_k \sin \bar{\theta}_k)^2 [1 + \cos 2(\phi_k
-\bar{\phi}_k)] \cr 
&+& 2 \sin \theta_k \sin \bar{\theta}_k \cos \theta_k \cos \bar{\theta}_k
\cos (\phi_k -\bar{\phi}_k)\}.
\end{eqnarray}
By integrations with respect to $\phi_k$ and
$\bar{\phi}_k$, we obtain  $\int\int d\phi_k d\bar{\phi}_k \cos
(\phi_k -\bar{\phi}_k) = \int\int d\phi_k 
d\bar{\phi}_k \cos 2(\phi_k -\bar{\phi}_k) = 0$, while $\int\int
\phi_k \bar{\phi}_k [\cos (\phi_k -\bar{\phi}_k)]^2 = \int\int d\phi_k
d\bar{\phi_k} [\cos 2(\phi_k -\bar{\phi}_k)]^2 = (2\pi)^2/2$.
  
Executing integrations in Eq.(\ref{eq:c23}) with respect to
$\phi_k$ and $\bar{\phi}_k$, and using the relations Eq.(\ref{eq:c7}) and
\begin{equation}
  \label{eq:c25}
e^{-i{\bf ky}} = e^{-iks\mu} = \sum_m (-i)^m (2m+1) j_m (ks)
P_m (-\mu) = \sum_m i^m (2m+1) j_m (ks) P_m (\mu),
\end{equation}
we can, therefore, reexpress Eq.(\ref{eq:c23}) as 
\begin{equation}
  \label{eq:c26}
\langle (\Theta_{pp})^2 \rangle = \sum_l  \sum_{l'}  \sum_m
\sum_{m'} \Bigl(A^I_{ll'mm'} + A^{II}_{ll'mm'} +
A^{III}_{ll'mm'}\Bigr), 
\end{equation}
where the terms $A^I_{ll'mm'}, A^{II}_{ll'mm'}$ and $A^{III}_{ll'mm'}$ 
come from the terms without $\phi_k$ and $\bar{\phi}_k$, the 
coefficients of $\cos^2 (\phi_k -\bar{\phi}_k)$, and the
coefficients of $\cos^2 2(\phi_k -\bar{\phi}_k)$, and their lengthy
expressions are shown in Appendix A.

Moreover let us replace the terms of $\mu j_l P_l, \ \mu^2 j_l P_l$
and $\mu^3 j_l P_l$ by $j_l^{(1)} P_l, \ j_l^{(2)} P_l$ and $j_l^{(3)}
P_l$ using Eqs. (\ref{eq:c16}) and  
(\ref{eq:c17}) and execute integrations with respect to $\mu$ and
$\bar{\mu}$ using Eq.(\ref{eq:c11a}). Then we obtain 
\begin{equation}
  \label{eq:c27}
\langle (\Theta_{pp})^2 \rangle =  \sum_l \sum_{l'}
\Bigl(B^I_{ll'} +B^{II}_{ll'}+B^{III}_{ll'}\Bigr),  
\end{equation}
where 

\begin{eqnarray}
  \label{eq:c29}
B^I_{ll'} &=& {1\over 8}(2\pi)^{-2} (2l+1) (2l'+1)
 \int\int dk d\bar{k}
(k\bar{k})^2 {\cal P}_F(k){\cal P}_F(\bar{k}) \cr  
&\times& \Bigl\{(k\bar{k})^2 \int^{\lambda_e}_{\lambda_o}
\int^{\lambda_e}_{\lambda_o} d\lambda d\bar{\lambda} 
P'(\lambda) 
P'(\bar{\lambda}) j_l^{(2)}(kr) j_{l'}^{(2)}(\bar{k}\bar{r}) 
+ 2 (k\bar{k})^2 \int^{\lambda_e}_{\lambda_o} d\lambda P''(\lambda) 
\int^{\lambda}_{\lambda_o} d\bar{\lambda} P(\bar{\lambda})j_l^{(2)}(kr)
j_{l'}^{(2)}(\bar{k}\bar{r}) \cr
&+& \int^{\lambda_e}_{\lambda_o} d\lambda \Bigl[P'(\lambda)
\Bigl({5\over 2}k\bar{k} j_l^{(1)}(kr) 
j_{l'}^{(1)}(\bar{k}r) + k^2 j_l^{(2)}(kr)
j_{l'} (\bar{k}r) + \bar{k}^2 j_l(kr) j_{l'}^{(2)} (\bar{k}r)\Bigr) \cr
&+& {3\over 14}P(\lambda)P'(\lambda) (k\bar{k})^2 \Bigl(
 j_l(kr) j_{l'}(\bar{k}r) -3j_l^{(2)}(kr) 
j_{l'}(\bar{k}r) -3j_l(kr) j_{l'}^{(2)}(\bar{k}r) \cr
&+& {1\over 3} j_l^{(2)}(kr) j_{l'}^{(2)}(\bar{k}r) \Bigr) \Bigr] 
\Bigr\}^2, 
\end{eqnarray}
\begin{eqnarray}
  \label{eq:c30}
B^{II}_{ll'} &=& {1\over 32}(2\pi)^{-2}  (2l+1) (2l'+1)
 \int\int dk d\bar{k} (k\bar{k})^4 {\cal P}_F(k){\cal P}_F(\bar{k})
\cr  
&\times& \int^{\lambda_e}_{\lambda_o} d\lambda P'(\lambda)
 \Bigl[\Bigl(j_l(kr) -j_l^{(2)}(kr) \Bigr) 
-{26\over 7} k\bar{k} P(\lambda)  \Bigl(j_l^{(1)}(kr)
-j_l^{(3)}(kr) \Bigr) \Bigl] j_{l'}(kr) \cr
&\times& \int^{\lambda_e}_{\lambda_o} d\eta P'(\eta)
\Bigl[ \Bigl(j_{l'}(\bar{k}s) 
-j_{l'}^{(2)}(\bar{k}s) \Bigr) -{26\over 7} k\bar{k} P(\eta)
 \Bigl(j_{l'}^{(1)}(\bar{k}s) -j_{l'}^{(3)}(\bar{k}s) \Bigr) \Bigl]
j_l (\bar{k}s),   
\end{eqnarray}
\begin{eqnarray}
  \label{eq:c31}
B^{III}_{ll'} &=& 2 \Bigl({3\over 56}\Bigr)^2(2\pi)^{-2}
 (2l+1) (2l'+1)  \int\int dk d\bar{k} (k\bar{k})^6 {\cal
P}_F(k){\cal P}_F(\bar{k}) \cr  
&\times& \Big\{\int^{\lambda_e}_{\lambda_o} d\lambda
P(\lambda)P'(\lambda) \Bigl[j_l(kr) j_{l'}(\bar{k}r) - 
j_l^{(2)} (kr) j_{l'}(\bar{k}r) -j_l(kr) j_{l'}^{(2)} (\bar{k}r)
+j_l^{(2)}(kr) j_{l'}^{(2)} (\bar{k}r) \Bigr] \Bigr\}^2,
\end{eqnarray}
where $r=\lambda_o -\lambda,\ \bar{r} = \lambda_o -\bar{\lambda},\ s
= \eta_o - \eta, \ \eta_e =
\lambda_e$, \ and \ $\eta_o = \lambda_o$.  

Next let us consider two directions with unit directional vectors
${\bf e}_1$ and ${\bf e}_1$. If we define $\mu_1, \mu_2$ and $\beta$
as $ \mu_1 = {\bf k}{\bf e}_1/k,\  \mu_2 = {\bf k}{\bf e}_2/k$ \
and $\cos \beta = {\bf e}_1 {\bf e}_2$, respectively, we have a
mathematical relation  
\begin{equation}
  \label{eq:c36}
\int \int d \phi_k d\theta_k \sin \theta_k P_l (\mu_1) P_l (\mu_2) =
{4\pi \over 2l+1} P_l (\cos \beta).  
\end{equation}
Using Eq.(\ref{eq:c36}) it is found that the correlation between
$\Theta_{pp}$'s in the two directions is expressed as
\begin{equation}
  \label{eq:c37}
 \langle \Theta_{pp}({\bf e}_1) \Theta_{pp}({\bf e}_2) \rangle
= {1\over 2} \sum_l \sum_{l'}  \Bigl(B^I_{ll'} +B^{II}_{ll'}
+B^{III}_{ll'}\Bigr) P_l (\cos \beta) P_{l'} (\cos \beta).  
\end{equation}
Here let us expand $P_l P_{l'}$ by $P_n$ as
\begin{equation}
  \label{eq:c38}
P_l(\cos \beta) P_{l'}(\cos \beta) = \sum_n b_{ll'n} P_n(\cos \beta).
\end{equation}
The derivation of the coefficient $b_{ll'n}$ is shown in Appendix B. 
Then the correlation reduces to
\begin{equation}
  \label{eq:c39}
(T_0)^2 \langle \Theta_{pp}({\bf e}_1) \Theta_{pp}({\bf e}_2) \rangle
= \sum_n {2n+1 \over 4\pi} C_n^{(2)} P_n(\cos \beta),
\end{equation}
where
\begin{equation}
  \label{eq:c40}
C_n^{(2)} = {4\pi \over 2n+1}(T_0)^2 \sum_{l,l'}
\Bigl(B^I_{ll'} +B^{II}_{ll'} +B^{III}_{ll'} \Bigr) b_{ll'n}.
\end{equation}
The expressions for $\langle \mathop{\delta}_2 T/T \rangle, \ \langle
(\Theta_{pp})^2 \rangle$ and $\langle 
\Theta_{pp}({\bf e}_1) \Theta_{pp}({\bf e}_2) \rangle$ in
Eqs.(\ref{eq:c14}), (\ref{eq:c27}) and (\ref{eq:c39}) are our new
result  which will 
be useful to derive the second-order power spectra.

\section{Concluding remarks}
In this paper we derived the average value $\langle \mathop{\delta}_2 
T/T \rangle$ and the second-order power spectra $C_n^{(2)}$ of CMB
anisotropies due to primordial random density perturbations, which
include two random variables $\alpha ({\bf k})$ and $\alpha ({\bf
\bar{k}})$, using the average values of the products of $\alpha 
({\bf k})$. The average value $\langle \mathop{\delta}_2 
T/T \rangle$ does not vanish and has no angular 
dependence. It should be regarded as the monopole component of
temperature fluctuations and the second-order
angular correlation is described by the power spectra $C_n^{(2)}$. 
Since we have not derived $\langle \mathop{\delta}_1 T/T
\mathop{\delta}_3 T/T \rangle$ yet, our analysis of second-order power
spectra is incomplete. But we think $\langle \Theta_{pp}^2 \rangle$ may
represent the essential feature of second-order power spectra, that is,
their $l$-dependence. In the next step we will analyze $\langle 
\mathop{\delta}_1 T/T \mathop{\delta}_3 T/T \rangle$ with the 
third-order metric perturbations for completeness.

In the case $\Lambda = 0$, the first-order temperature fluctuations
due to the integrated Sachs-Wolfe effect vanish, while the second-order
ones do not vanish and play a dominant role in the integrated Sachs-Wolfe
effect, which should not be neglected. In the cosmological local void
model (LVM)\cite{lvm,cele,aln,bis}, the exterior region is 
described in terms of the Einstein-de Sitter model, and so the main
part of the integrated Sachs-Wolfe effect in LVM is of second-order.  
In the interior region we use open low-density models with 
$\Omega_0 < 1$, in which we have nonzero first-order Sachs-Wolfe 
effect. It is important for showing the observational reality of LVM 
to derive them.  

In the case $\Lambda \ne 0$, the first-order temperature fluctuations
due to the integrated Sachs-Wolfe effect decrease rapidly with the
increase of the redshift $z$, but the second-order temperature
fluctuations due to the 
integrated Sach-Wolfe effect decrease more slowly, and so the latter
fluctuations may be dominant over the first-order fluctuations at the
early stage. This situation is explained in a separate paper\cite{ti} 
using a simple model of density perturbations. Thus a characteristic
behavior of CMB power spectra due to second-order temperature
fluctuations will be measured 
through the future precise observation and bring useful informations
on the structure and evolution of our universe in the future. 

\begin{acknowledgments}
The author thanks K.T. Inoue and referees for helpful discussions and
comments. 
\end{acknowledgments}

\appendix
\section{First expressions of $A^I_{ll'mm'}, A^{II}_{ll'mm'}$ and
$A^{III}_{ll'mm'}$} 
\begin{eqnarray}
  \label{eq:c50}
A^I_{ll'mm'} &=& {2\over 8^2}(2\pi)^{-2} (-i)^{(l+l')} i^{(m+m')}
(2l+1) (2l'+1) (2m+1) (2m'+1) \int\int dk d\bar{k} (k\bar{k})^2
 d\mu d\bar{\mu} {\cal P}_F (k){\cal P}_F (\bar{k}) \cr
&\times& \int^{\lambda_e}_{\lambda_o} d\lambda \Bigl\{
\int^{\lambda_e}_{\lambda_o} d\bar{\lambda} P'(\lambda) 
P'(\bar{\lambda})  (k\bar{k})^2 (\mu\bar{\mu})^2 j_l(kr)
j_{l'}(\bar{k}\bar{r}) P_l(\mu) P_{l'}(\bar{\mu}) 
+ \Bigl[P'(\lambda) ({5\over 2}k\bar{k}\mu\bar{\mu}
+k^2\mu^2 + \bar{k}^2\bar{\mu}^2) \cr
&+&{1\over 7} P(\lambda)P'(\lambda)
(k\bar{k})^2 \Bigl(2 (\mu\bar{\mu})^2 -6(\mu^2 +\bar{\mu}^2) 
+3 -{3\over 2}(1-\mu^2)(1-\bar{\mu}^2)\Bigr)\Bigr] \cr 
&\times& j_l(kr)j_{l'}(\bar{k}r)P_l(\mu) P_{l'}(\bar{\mu})  
+ 2 P''(\lambda) \int^\lambda_{\lambda_o} d\bar{\lambda}
P(\bar{\lambda}) (k\bar{k}\mu\bar{\mu})^2 j_l(kr)
j_{l'}(\bar{k}\bar{r})P_l(\mu) P_{l'}(\bar{\mu}) \Bigr\}\cr 
&\times& \int^{\eta_e}_{\eta_o} d\eta \Bigl\{
\int^{\eta_e}_{\eta_o} d\bar{\eta} P'(\eta) 
P'(\bar{\eta})  (k\bar{k})^2 (\mu\bar{\mu})^2 j_m(k s)
j_{m'}(\bar{k} \bar{s}) P_m(\mu) P_{m'}(\bar{\mu}) 
+ \Bigl[P'(\eta) ({5\over 2}k\bar{k}\mu\bar{\mu}
+k^2\mu^2 + \bar{k}^2\bar{\mu}^2) \cr
&+&{1\over 7} P(\eta)P'(\eta)
(k\bar{k})^2 \Bigl( 2 (\mu\bar{\mu})^2 -6(\mu^2 +\bar{\mu}^2)
+3 -{3\over 2}(1-\mu^2)(1-\bar{\mu}^2)\Bigr)\Bigr]\cr 
&\times& j_m(k s)j_{m'}(\bar{k} s)P_m(\mu)P_{m'}(\bar{\mu}) 
+ 2 P''(\eta) \int^{\eta}_{\eta_o} d\bar{\eta} P(\bar{\eta})
(k\bar{k}\mu\bar{\mu})^2 j_m(k s) j_{m'}(\bar{k} \bar{s})P_m(\mu)
P_{m'}(\bar{\mu}) \Bigr\}, 
\end{eqnarray}
\begin{eqnarray}
  \label{eq:c51}
A^{II}_{ll'mm'} &=& {2\over 16^2}(2\pi)^{-2} 
(-i)^{(l+l')} i^{(m+m')} (2l+1) (2l'+1) (2m+1)
(2m'+1) \int\int dk d\bar{k} (k\bar{k})^4
 d\mu d\bar{\mu} {\cal P}_F(k){\cal P}_F(\bar{k}) \cr
&\times&  \int^{\lambda_e}_{\lambda_o} d\lambda 
P'(\lambda) \Bigl[1 - {26\over 7} P(\lambda) k\bar{k} \mu\bar{\mu}
\Bigr] j_l(kr) 
j_{l'}(\bar{k}r) (1-\mu^2)P_l(\mu) P_{l'}(\bar{\mu})  \cr
&\times& \int^{\eta_e}_{\eta_o} d\eta  P'(\eta)\Bigl[1 -{26\over 7}
P(\eta) k\bar{k} \mu\bar{\mu} \Bigr] 
 j_m(k s) j_{m'}(\bar{k} s) (1-\bar{\mu}^2)P_m(\mu)P_{m'}(\bar{\mu}) , 
\end{eqnarray}
\begin{eqnarray}
  \label{eq:c52}
A^{III}_{ll'mm'} &=& 2 \Bigl({3\over 112}\Bigr)^2 (2\pi)^{-2}
(-i)^{(l+l')} i^{(m+m')} (2l+1) 
(2l'+1) (2m+1) (2m'+1) \int\int dk d\bar{k} (k\bar{k})^6
 d\mu d\bar{\mu} {\cal P}_F(k){\cal P}_F(\bar{k}) \cr
&\times&  \int^{\lambda_e}_{\lambda_o} d\lambda 
P(\lambda) P'(\lambda) j_l(kr)j_{l'}(\bar{k}r)
(1-\mu^2)(1-\bar{\mu}^2) P_l(\mu) P_{l'}(\bar{\mu})  \cr 
&\times& \int^{\eta_e}_{\eta_o} d\eta P(\eta)P'(\eta)
 j_m(k s) j_{m'}(\bar{k}s) (1-\mu^2)(1-\bar{\mu}^2)
P_m(\mu)P_{m'}(\bar{\mu}),  
\end{eqnarray}
where $r = \vert {\bf x} \vert, \ s = \vert {\bf y} \vert,
\ r=\lambda_o -\lambda,\ \bar{r} = \lambda_o -\bar{\lambda},\ s
= \eta_o - \eta, \ \bar{s} = \eta_o -\bar{\eta}, \ \eta_e =
\lambda_e$, \ and \ $\eta_o = \lambda_o$.

\section{Derivation of $b_{ll'n}$}

Since the Legendre function $P_m (z)$ is a finite power series of
$z^j$ with integers $j (\le m)$, $P_m(z) P_{m'}(z)$ can be also
expressed as a finite power series of $z^j$ with integers $r$
satisfying $0 \le j \le m+m'$. On the other hand, $z^n$ is expressed
as a finite series of $P_m (z)$ with $0 \le m \le n$, so that $P_m(z)
P_{m'}(z)$ can be expressed as a finite series of $P_n (z)$ with $0
\le n \le m+m'$. Their examples are
\begin{eqnarray}
  \label{eq:ac1}
P_1P_2 &=& {2\over 5}P_1 + {3\over 5}P_3, \cr
(P_2)^2 &=& {1\over 5}P_0 + {2\over 7}P_2 + {18\over 35}P_4, \cr
P_1P_3 &=& {3\over 7}P_2 + {4\over 7}P_4, \cr
P_2P_3 &=& {9\over 35}P_1 + {4\over 15}P_3 +{10\over 21}P_5. 
\end{eqnarray}
In these cases the coefficients $b_{ll'n}$ in Eq.(\ref{eq:c38}) are  
\begin{eqnarray}
  \label{eq:ac2}
b_{1 2 1} &=& 2/5, \qquad b_{1 2 3} = 3/5, \cr
 b_{220} &=& 1/5,\qquad b_{222} =2/7,\qquad b_{224} = 18/35, \cr
b_{132} &=& 3/7, \qquad b_{134} = 4/7, \cr
b_{231} &=& 9/35, \qquad b_{233} = 4/15, \qquad b_{235} =10/21.   
\end{eqnarray}
The general formula of these series are classified into the following
three cases.

\noindent (1) even-even type
\begin{eqnarray}
  \label{eq:ac3}
P_{2m} P_{2m'} &=& \sum^{m+m'}_{s=0} b_{2m\ 2m'\ 2s} \ P_{2s}, \cr
b_{2m\ 2m'\ 2s} &=& {4s+1 \over 2^{2m+2m'}} \sum^m_{j=0}
\sum^{m'}_{j=0}  (-1)^{m+m'+j+j'} (2j+2j')!(2m+2j)! (2m'+2j')! \cr
&/& \Bigl[ (m+j)! (m-j)! (2j)! (m'+j')! 
(m'-j')! (2j')! (2j+2j'-2s)!! (2j+2j'+2s+1)!! \Bigr],
\end{eqnarray}
\noindent (2) odd-odd type
\begin{eqnarray}
  \label{eq:ac4}
P_{2m+1} P_{2m'+1} &=& \sum^{m+m'+1}_{s=0} b_{2m+1\ 2m'+1\ 2s} \ P_{2s},
\cr  
b_{2m+1\ 2m'+1\ 2s} &=& {4s+1 \over 2^{2m+2m'+2}} \sum^m_{j=0}
\sum^{m'}_{j'=0} (-1)^{m+m'+j+j'}(2j+2j'+2)! (2m+2j+2)! (2m'+2j'+2)! \cr
&/& \Bigl[(m+j+1)! (m-j)! (2j+1)! (m'+j'+1)! (m'-j')! (2j'+1)! \cr
&\times&(2j+2j'-2s+2)!! (2j+2j'+2s+3)!! \Bigr],
\end{eqnarray}
\noindent (3) even-odd type
\begin{eqnarray}
  \label{eq:ac5}
P_{2m} P_{2m'+1} &=& \sum^{m+m'}_{s=0} b_{2m\ 2m'+1\ 2s+1} \ P_{2s+1},
\cr 
b_{2m\ 2m'+1\ 2s+1} &=& {4s+3 \over 2^{2m+2m'+1}} \sum^m_{j=0}
\sum^{m'}_{j'=0} (-1)^{m+m'+j+j'} (2j+2j'+1)! (2m+2j)! (2m'+2j'+2)! \cr
&/& \Bigl[(m+j)! (m-j)! (2j)! (m'+j'+1)! (m'-j')! (2j'+1)! \cr
&\times&(2j+2j'-2s)!! (2j+2j'+2s+3)!! \Bigr],
\end{eqnarray}
where the rules of $!$ and $!!$ are
\begin{eqnarray}
  \label{eq:ac6} 
n! &=& 1\cdot 2\cdot \cdots (n-1)\cdot n \qquad {\rm for} \qquad n \geq 1, \cr
0! &=& 1, \cr
n! &=& \infty \qquad {\rm for} \qquad n < 0, 
\end{eqnarray}
and 
\begin{eqnarray}
  \label{eq:ac7} 
(2n)!! &=& 2\cdot 4\cdot \cdots (2n-2)\cdot 2n \qquad {\rm for} \qquad n \geq 1,
\cr 
(2n+1)!! &=& 1\cdot 3\cdot \cdots (2n-1)\cdot 2n+1 \qquad {\rm for} \qquad n \geq
0, \cr 
0!! &=& 1,  \cr
n!! &=& \infty \qquad {\rm for} \qquad n < 0. 
\end{eqnarray}
%

%



\end{document}